# Factor Analysis by R Programming to Assess Variability Among Environmental Determinants of the Mariana Trench

## Polina LEMENKOVA[1,*]

[1]*Ocean University of China, College of Marine Geoscience, 238 Songling Road, Laoshan, 266100, Qingdao, China*

## ABSTRACT

The aim of this work is to identify main impact factors affecting variations in the geomorphology of the Mariana Trench which is the deepest place of the Earth, located in the west Pacific Ocean: steepness angle and structure of the sediment compression.

The Mariana Trench presents a complex ecosystem with highly interconnected factors: geology (sediment thickness and tectonics including four plates that Mariana trench crosses: Philippine, Pacific, Mariana, Caroline), bathymetry (coordinates, slope angle, depth values in the observation points). To study such a complex system, an objective method combining various approaches (statistics, R, GIS, descriptive analysis and graphical plotting) was performed.

Methodology of the research includes following clusters: R programming language for writing codes, statistical analysis, mathematical algorithms for data processing, analysis and visualizing diagrams, GIS for digitizing bathymetric profiles and spatial analysis. The statistical analysis of the data taken from the bathymetric profiles was applied to environmental factors, i.e. coordinates, depths, geological properties sediment thickness, slope angles, etc. Finally, factor analysis was performed by R libraries to analyze impact factors of the Mariana Trench ecosystem. Euler-Venn logical diagrams highlighted similarities between four tectonic plates and environmental factors.

The results revealed distinct correlations between the environmental factors (sediment thickness, slope steepness, depth values by observation points, geographic location of the profiles) affecting Mariana Trench morphology.

The research demonstrated that coding on R language provides a powerful and highly effective statistical tools, mathematical algorithms of factor analysis to study ocean trench formation.

*Corresponding Author
E-mail: pauline.lemenkova@gmail.com

# 1. INTRODUCTION

This research focuses on the geomorphological formation factors of the Mariana Trench (Fig. 1), a long and narrow topographic depressions of the sea floor located in the west Pacific Ocean. Mariana Trench is the deepest part of the ocean floor, a distinctive morphological feature of the convergent plate boundaries, along which lithospheric plates move towards each other, located about 200 km parallel to a volcanic island arc (Karato *et al.*, 2001).

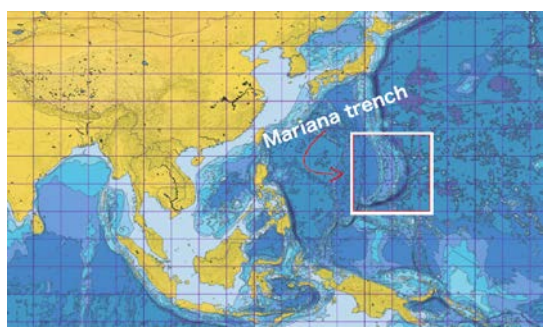

**Figure 1.** Study area: Mariana Trench

A trench marks the position at which the subducting slab descends beneath another lithospheric slab (Deschamps and Lallemand, 2003).

## 1.1. Geology: Sedimentation

Thin Quaternary clayey sediments overlie a 400 m thick alternating sequence of the Early Cretaceous quartz-trachyte pillow lava and Early to Late Cretaceous radiolarian cherts in the oceanward slope of the Mariana Trench (Hirano *et al.*, 2002). The sediment cover is less than 10 m thick and is composed of soft, bedded clay with sand-size ash fragments. This clayey sediment covers a very gentle terrace 500 m wide (Curtis and Moyer, 2005). The turbidite currents are being generated on the outer shelf and upper continental slope, due to the precipitation and transport via the submarine canyons to the axial part of the trench. The speed of such sedimentation in the bottoms of trenches closely depends on the volumes of the incoming sediments ranging from 300 to 3000 mm/thousand years (Ishizuka *et al.*, 2018). As a result of such sedimentation, over one kilometer of sediments have accumulated in the southern Aleutian and Chilean trenches over the last several hundred thousand years. On the bottom of the deep ocean trenches a longitudinal main channel is usually formed under the influence of turbidite currents (Pabst *et al.*, 2012). It creates favorable conditions for distinctive movements of the sand sediments that shape large elongated accumulations of the sediments (Ernst, 2001). As a hadal trench, Mariana Trench belongs to the one of the deepest 45% of the ocean's depth range, and its unique topography disrupts the continental shelf-slope-rise to abyssal plain continuum, resulting in an array of deep isolated habitats. The often abrupt and distinct topography is likely to further promote speciation through high hydrostatic pressure, remoteness from surface derived food sources and geographic isolation (Eustace *et al.*, 2016).

## 1.2. Geomorphology: Cracks

Mariana Trench is notable for its well-developed systems of cracks in the structure. Cracks of the Mariana Trench are abundantly developed at the edge of the terrace, in the middle and upper part of the terrace (Heuret and Lallemand, 2005). The explanation of the formation of these cracks is, besides gravitational instability, that horizontal extension is due to the trench-ward increase of the dip angle of the ocean floor surface, and that this might cause slope instability due to gravity pull. The cracks on the surface of the seabed on the oceanward slopes of the Mariana Trench are attributed to tensile rock failure induced by a combination of slope instability and earthquake shaking (Schellart, 2007). The cracks of the Mariana Trench are found on the horizontal or very gentle slopes just above steep cliffs, and are mostly elongated in directions nearly parallel to the strike of the cliffs, although some are aligned, branching or merging have no vertical nor lateral displacement, that could be explained by open tension fractures.

Surface edges of the Mariana Trench cracks are generally very sharp, indicating their young origin, with a noticeable pressure ridge caused by mud overflow during closure of one crack. The cracks were formed at the horizontally stretched surface of the down-going subducting oceanic plate under tensional stress. This tensional stress may have been caused by a combination of gravitational slope instability plus additional inertia during earthquake shaking which occurs close to these areas (DeMets *et al.*, 1990). Thus, the cracks were formed at topographically specific areas, where gravitational instability or instability occurs at the edges of the slope or top of the ridge (Ishizuka *et al., 2010).*

**1.3. Tectonics: Slab Movement**

Understanding the mechanisms of trench migration (retreat or advance) is crucial to characterizing the driving forces of Earth's tectonics plates, the origins of subducting slab morphologies in the deep mantle, and identifying the characteristics of subduction zones systems, which are among the fundamental issues of solid Earth science (Yoshida, 2017).

Recent studies revealed (Faccenna *et al.,* 2009) that Mariana Trench advance toward the upper plate corresponding to the subduction of very old, Mesozoic oceanic lithosphere. The Mariana Arc lavas are relatively enriched in Molybdenum (Mo) and have $\delta^{98/95}$ Mo significantly greater than MORB (apart from the samples from the island of Agrigan). This implicates the addition of a Mo-rich fluid with $\delta^{98/95}$Mo ~+0.05% to the mantle wedge beneath most islands (Freymuth *et al*, 2015). Recently updated analyses of global plate motions indicate that significant trench advance is also rare on Earth, being largely restricted to the Marianas–Izu–Bonin arc (Čížková and Bin, 2015). The effects of trench migration on the descent of subducted slabs are discussed by Griffiths *et al.*, 1995. The global plate tectonics movements shaping the seafloor bathymetry can be described as follows. The seafloor spreading creates an axial rift and corrugated hills. Spreading ridges are formed by nearby faults where the most destructive earthquakes occur. Subduction of the cooled plate into the mantle causes creation of the deep ocean trenches and here, as a consequence, the major earthquakes and tsunamis originate. The plates act as giant radiators of the heat cooling, thickening, and gradually subsiding by progressing from ridge to trench. It explains the appearance of the double seismic zone beneath the Mariana Island arc (Samowitz and Forsyth, 1981). In such a way, tectonic plates form the large-scale pattern of the system of ridges and deep ocean basins.

The deformation of the upper plates, as well as the surface migration of the ocean trenches of the Pacific Ocean are the prevailing observables of the dynamics of the inter parts of the plates, as described in various research papers (e.g., Miller *et al*., 2004). Therefore, the classification of the tectonic plate boundaries according to their cross-correlation with the kinematical and geometrical properties of the plate boundaries should reflect their geodynamics. Various research papers have discussed a problem of the slab movement around the trench (e.g. Fujioka *et al*., 2002; Funiciello *et al*., 2008; Heuret *et al.*, 2012.)

## 2. MATERIAL AND METHOD

### 2.1. GIS Data Processing

The GIS part of the research is performed in the QGIS by creating 25 bathymetric profiles crossing Mariana Trench. Each profile has a length of 1000 km, a distance gap of 100 km in between and 518 measurement points in each. The total dataset comprises of large amount of 12950 observations.

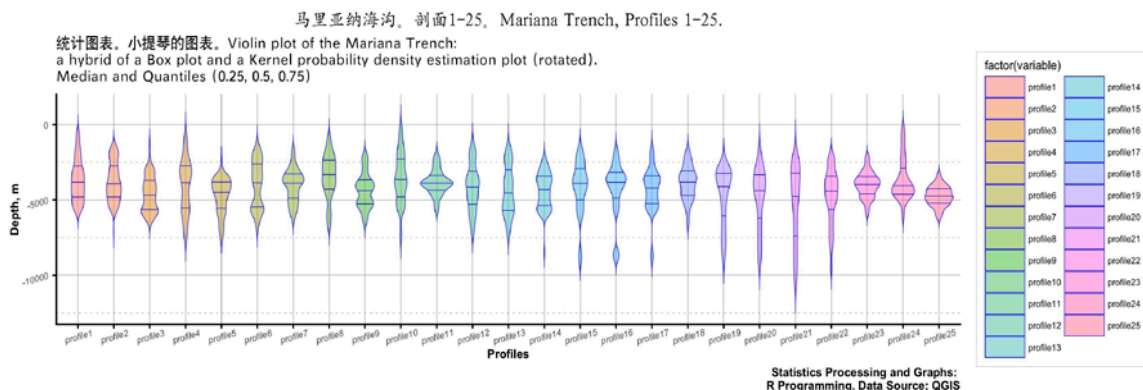

**Figure 2.** Violin plots showing depth data distribution for observation points of the 25 bathymetric profiles of the Mariana Trench

Data includes tectonic plates, topography of the profiles, slope angle, depths, aspect class, location of the igneous volcanic areas, sediment thickness. In the QGIS several tasks were performed using thematic plugins (e.g. coordinate re-projecting). The geospatial data have been uploaded to form the GIS project (bathymetric features, sediment thickness, location of the igneous volcanic zones). Multiple thematic layers were upload into the GIS system that include among others marine geology data stores in layers, settings, coordinate system, parameters, etc. The geometry of the profiles has been digitized along the Mariana Trench, the attributes (profiles names and coordinates) have been entered. Every profile had a length of 1.000 km and the distance between each two neighbor profiles 100 km along the path of the trench. The GIS processing resulted in bathymetric 25 profiles. The GIS project has been re-projected into the UTM cartesian coordinate system (square N-55).

### 2.2. Analysis of Data Distribution

The violin plots (Fig. 2) were created to show Kernel probability density distribution of the bathymetric observations, as multimodal distributions with multiple peaks. Kernel density distribution plot as shown on the Fig. 2 was created using library {violinmplot} of R in a combined plot, which includes calculated quantiles for 0.25 and 0.75 of the data pool. The kernel density estimation (KDE) is a non-parametric way to estimate the probability density function of a random variable of the observation depth. The KDE has been performed by smoothing data points across the sample points in the profiles 1:25 of the Mariana Trench. Kernel density estimates can be endowed with properties such as smoothness or continuity of the bathymetric data by using a suitable kernel. Technically, besides {violinplot} library, a kernel density estimation function can be implemented in R through the density, the {bkde} function in the {KernSmooth} library, as well as through the {kde} function in the {ks} library. To use the {kde.R} function, it is not required to install any packages or libraries, while {violinplot} package was installed and activated prior to the current work.

## 2.3. Factor Analysis

To perform factor analysis and principal component analysis, an approach based on the use of R scripting was used (Fig. 3). To this end the R libraries for scripting simulations were used aimed to model principal factors that affect the morphology of the trench.

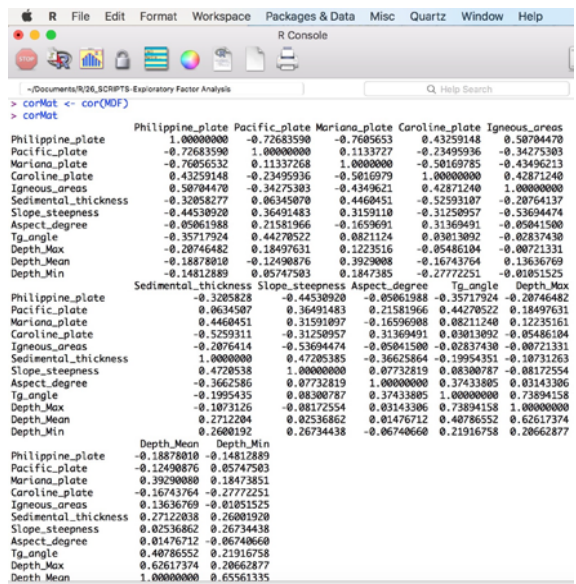

**Figure 3.** A programming script for factor analysis written on R code.

For the factor analysis the entire modeling process including grid generation, model setup, execution and analysis of the results with visualized correlation matrix showing the more important factors impacts according to their values, has been carried out from a single R script, with the results of the correlation matrix shown on Fig. 4. The standardized loadings based upon correlation matrix of factor analysis of the Mariana Trench (Fig. 5) show the impact values of each factor. For this research the fa() function from R was used (Fig. 3), which received the following primary arguments: r: the correlation matrix; Nfactors: number of factors to be extracted; rotate: one of several matrix rotation methods, such as "varimax" or "oblimin" (Fig. 8).

**Figure 4.** Correlation matrix of factors of the Mariana Trench. R Programming

Next, an alternative to factor for the components analysis was performed using {FactoMiner} library by cluster analysis iCLUST (Fig. 9). This goal is the same as factor or components analysis, but methodologically, it reduces the complexity of the data and attempts to identify homogeneous sub-groups. Here, the exploratory factor analysis aims to extract a more regular impact factors of the geologic morphology development, whereas cluster analysis and hierarchical dendrogram extract the groups and classes in a total cloud of the bathymetric observation points.

Finally, the factor analysis results in the illustrating cross-influences of the most notable factors: geological, geomorphological, tectonic, geographic and environmental parameters.

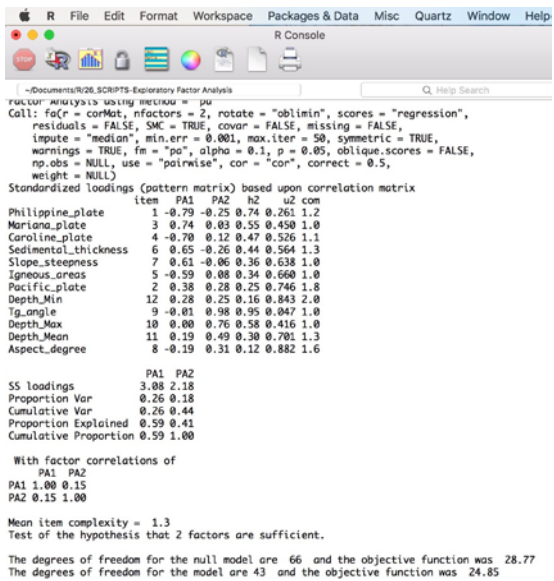

**Figure 5.** Standardized loadings based upon correlation matrix of factor analysis

This enables to quantitatively analyze correlation between the actual shape of the trench and its environmental impact factors (Fig. 6). Fm: one of several factoring methods, such as "pa" (principal axis) or "ml" (maximum likelihood), as shown on the Fig. 3.

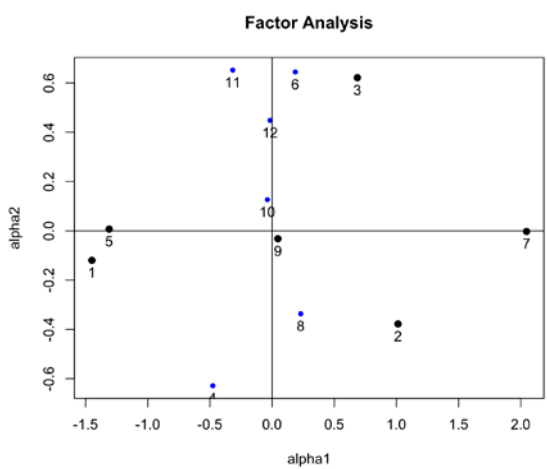

**Figure 6.** Exploratory factor analysis

The omega testing was done to find out two alternative estimates of the reliability that take into account the hierarchical structure of the inventory data are ω (Fig. 10 and Fig. 11). These were called using the omega function for factor analysis, R. The computed coefficient omega (hierarchical) (ωh) is an estimate of the general factor saturation of the performed test. Various factoring methods have been tested in this research to better describe data factors. Nevertheless, the fa() function provided the best results used for common factoring (Fig. 7). In this research the oblique rotation (rotate = "oblimin") has been used, which recognizes that there is likely to be some correlation between geomorphological factors affecting Mariana Trench formation. The principal axis factoring (fm = "pa") was used in this research, as the identifying the underlying constructs in the data was not necessary in this case.

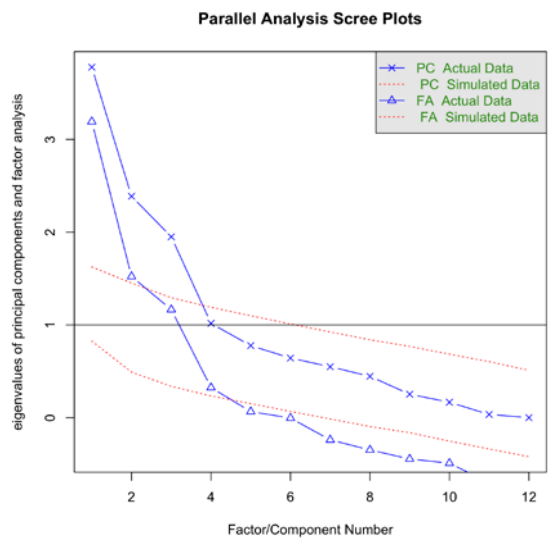

**Figure 7.** Parallel analysis scree plots

The exploratory factor analysis (EFA) matrix is shown on Fig. 8. Finally, the Euler-Venn plot (Fig. 12) has been drawn to visualize all possible crossings between the variables using {venn} library of R calling following script:

x <-list(Philippine = MDF$plate_phill,
Pacific = MDF$plate_pacif, Mariana = MDF$plate_maria,
Caroline = MDF$plate_carol,
Aspect = MDF$aspect_class,
Morphology = MDF$morph_class,

Slope = MDF$slope_class)
venn(x, ilabels = TRUE, col = "navyblue",
zcolor = "style")

## 3. RESULTS

Once all the cross-section profiles had
been inspected, the assessment of the
bathymetric observation points and other
environmental parameters (depth,
sediment thickness) has been met by the
factor analysis correlation matrix (Fig. 8).

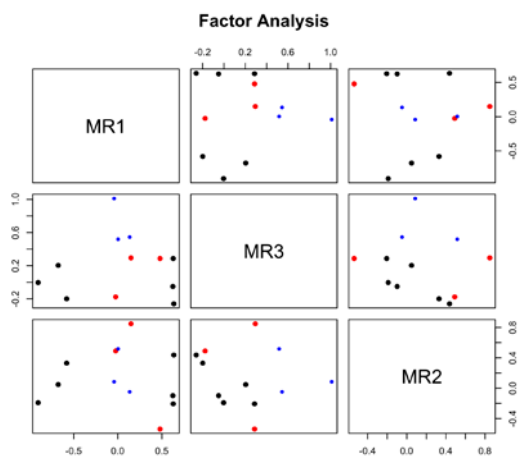

**Figure 8.** EFA correlation matrix.

The factor loadings (Fig. 5) enable to
assess the results of the factor solutions.
Thus, we can see that the sediment
thickness (factor solution = 0.65), slope
angle steepness (factor solution = 0.61),
and Mariana Plate location (factor solution
= 0.74) have high factor loadings >0.5
around 0.7 on the first factor (PA1).
Therefore, these factors are the most
influencing and representative for the
morphology of the trench. The sediment
thickness is mostly impacted by the slope
steepness degree. Two geophysical
indicators were particularly tested in the
comparative analysis (Fig. 5). It was
furthermore found that slope degree and
amplitude has important impact on the
sediment thickness, while the aspect
degree has lesser effect.

Secondly, the second column (PA2)
reveals (Fig. 5) that calculated tg° slope
angle has a factor solution = 0.98
(extremely high), following by depth
distribution (factor solution = 0.76).

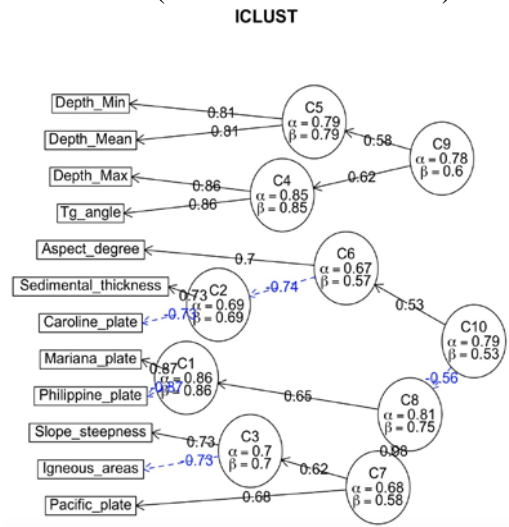

**Figure 9.** iCLUST analysis

The location of the Caroline and Philippine
Plates do not affect that much Mariana
Trench having negative factor solutions -
0.79 and -0.70, respectively), as well as a
much lower loading on PA2 (factor
solutions = 0.25 and 0.12) and that it has a
slight loading on factor PA1.

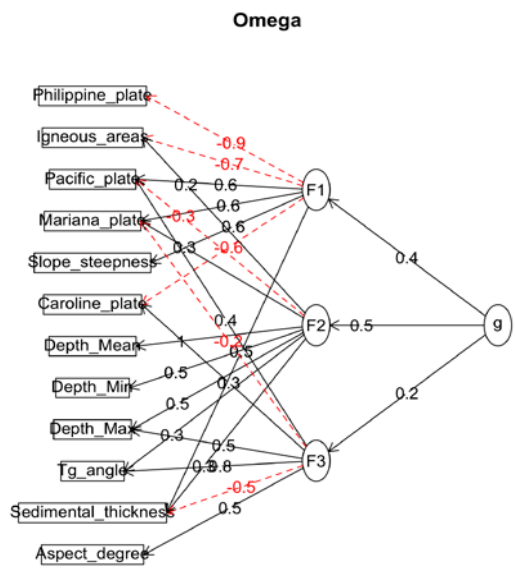

**Figure 10.** General factor saturation (ωh)

This suggests that statistics is less related to the concept of Caroline and Philippine Plates than Mariana Plate and its geomorphic environmental settings, such as tg° slope angle, sediment thickness, slope angle steepness.

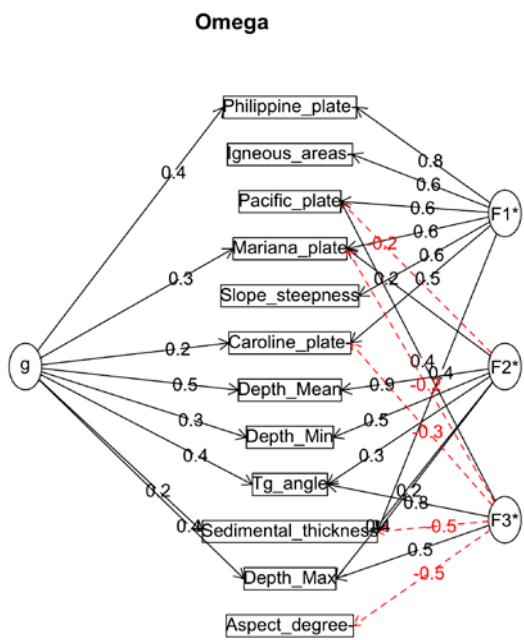

**Figure 11.** General factor saturation through omega coefficient: 3 factors (ω)

Furthermore, on the resulting table (Fig. 5) one can see that each factor accounted for the 20% (0.26% and 0.18%) of the variance in responses, leading to a factor solution that accounted for 100% of the total variance (Cumulative proportion PA2: 1.00) in the Mariana Trench morphology formation.

Finally, these factors are correlated in-between at 0.15 and recall that this choice of the oblique rotation allowed for the recognition of this relationship. The hypothesis testing was sufficient, with following factors having the highest score: tg° slope angle, sediment thickness, Mariana Plate location.

## 4. DISCUSSIONS

Current studies have revealed that there are factors influencing Mariana Trench geomorphic structure the most, namely: sediment thickness of the basement, slope angle steepness degree, angle aspect, bathymetric factors, such as depth at basement, means, median and minimal values, closeness of the igneous volcanic areas causing possible earthquakes, and geographic location across four tectonic continental plates including Mariana, Pacific, Philippine and Caroline.

The bathymetry of the ocean floor reflects plate tectonics processes, including trench movement, deformation and bending which is associated with mantle convection at the global scale. Therefore, studying combination of these factors is a prerequisite for the correct understanding of the complex processes that take place in the abyssal environment of the Mariana Trench. The tectonic plates, sediment thickness and location of the igneous volcanic areas around the cross-section profiles play an essential role in the morphology of the trench. Complex distribution of various environmental factors on the adjacent abyssal plains of the ocean contributes to the formation of the geomorphic features of the ocean bottom in the Mariana Trench.

## 5. CONCLUSIONS

The main innovative idea of this research was integrated usage of R programming language and statistical analysis towards marine geological studies. Geological studies of such complex geomorphic-oceanological system as Mariana Trench has specific points that distinguish it from the study of the valleys located on continents.

First, the rocks of the ocean floor are a closed object, which it is studied mainly by geophysical, i.e. remote, methods.

These observations include the study of cores obtained from drilling vessels or platforms, samples collected while dredging the seabed.

Secondly, the knowledge of the geology of the ocean is connected with a thorough understanding of the basics of geophysics, geography, geomorphology, oceanography, cartography, informatics and principles of the data interpretation.

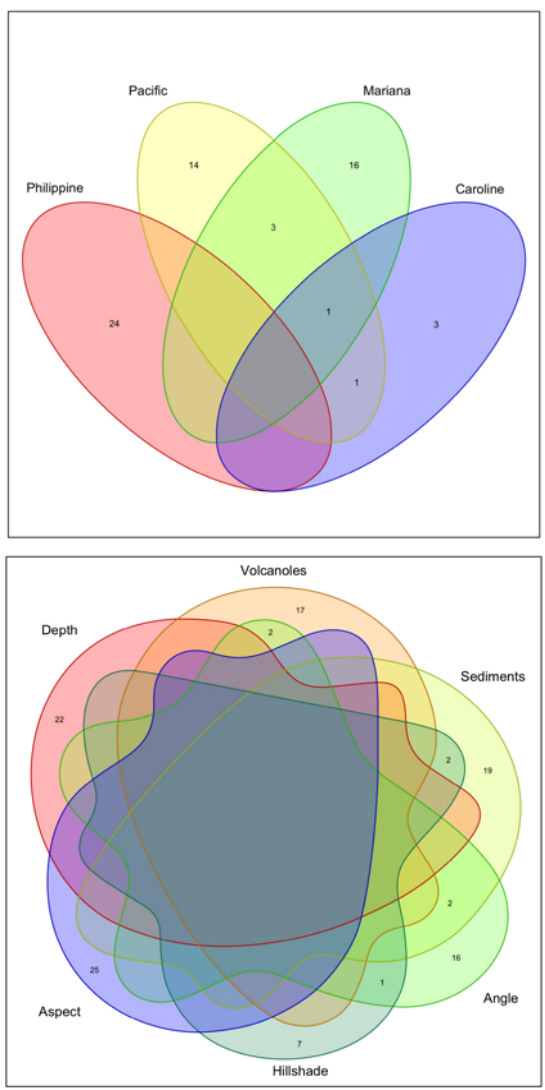

**Figure 12.** Euler-Venn logical diagram on correlation of impact factors affecting Mariana Trench. Upper: four tectonic plates. Below: environmental factors.

Therefore, a complex variety of methods is necessary to study Mariana Trench that has been demonstrated in this research by R programming language.

Current research was intended to highlight the problem of the Mariana Trench very complex formation consisting of a variety of environmental factors. Mariana Trench is formed as an ocean seafloor geomorphological structure located in the zones of the continental margin tectonic plates bending. Moreover, a special attention was paid to the application of the algorithms of factor analysis: scripts with screen shots of codes, correlation matrix, as well as visualization of factor analysis. As a case study of this research, the application of R programming towards geoscience studies successfully revealed impact factors affecting the abyssal morphology of the Mariana Trench. It contributed towards question of how we can measure and analyze the structure of trench in the least reachable location of the World Ocean.

**ACKNOWLEDGEMENTS**

The funding has been provided by the CSC, SOA, Marine Scholarship of China, Beijing [Grant #2016SOA002, 2016].

**6. REFERENCES**

Karato, S., Riedel, M.R., Yuen, D.A., (2001). Rheological structure and deformation of subducted slabs in the mantle transition zone: implications for mantle circulation and deep earthquakes. *Physics of the Earth and Planetary Interiors* 127: 1–7.

Deschamps, A. & Lallemand, S. (2003). Geodynamic setting of Izu-Bonin-Mariana boninites. In: Larter, R.D., Leat, P.T. (eds) (2003). Intra-Oceanic Subduction Systems: Tectonic and Magmatic Processes. Geological Society, London, Special Publications 219: 163-185.

Hirano, N., Ogawa, Y., Saito, K., (2002). Long-lived early Cretaceous seamount volcanism in the Mariana Trench, Western Pacific Ocean. *Marine Geology* 189: 371-379.

Curtis, A.C., Moyer, C.L., (2005). Mariana forearc serpentine mud volcanoes harbor novel communities of extremophilic Archaea[J]. *Geomicrobiology Journal* 30(5): 430-441.

Ishizuka, O., Hickey-Vargas, R., Arculus, R.J., Yogodzinski, G.M., Savov, I.P., Kusano, Y., McCarthy, A., Brandl, Ph., Sudo, M., (2018). Age of Izu–Bonin–Mariana arc basement. *Earth and Planetary Science Letters* 481: 80–90.

Pabst, S., Zack, Th., Savov, I.P., Ludwig, Th., Rost, D., Tonarini, S., Vicenzi, E.P., (2012). The fate of subducted oceanic slabs in the shallow mantle: Insights from boron isotopes and light element composition of metasomatized blueschists from the Mariana forearc. *Lithos* 132: 162–179.

Ernst, W.G., (2001). Subduction, ultrahigh-pressure metamorphism, and regurgitation of buoyant crustal slices — implications for arcs and continental growth, *Physics of the Earth and Planetary Interiors* 127(1–4): 253-275.

Eustace, R.M., Ritchie, H., Kilgallen, N.M., Piertney, S.B., Jamieson, A.J., (2016). Morphological and ontogenetic stratification of abyssal and hadal Eurythenes gryllus sensu lato (Amphipoda: Lysianassoidea) from the Peru–Chile Trench. *Deep-Sea Research I* 109: 91–98.

Heuret, A., Lallemand, S., (2005). Plate motions, slab dynamics and back-arc deformation. *Physics of the Earth and Planetary Interiors* 149: 31–51.

Schellart, W.P., (2007). The potential influence of subduction zone polarity on overriding plate deformation, trench migration and slab dip angle. *Tectonophysics* 445: 363–372.

DeMets, C., Gordon, R.G., Argus, D.F., Stein, S., (1990). Current plate motions, *Geophysical Journal International* 101: 425-478.

Ishizuka, O., Yuasa, M., Tamura, Y., Shukuno, H., Stern, R.J., Naka, J., Joshima, M., Taylor, R.N., (2010). Migrating shoshonitic magmatism tracks Izu–Bonin–Mariana intra-oceanic arc rift propagation. *Earth and Planetary Science Letters* 294: 111–122.

Yoshida, M., (2017). Trench dynamics: Effects of dynamically migrating trench on subducting slab morphology and characteristics of subduction zones systems. *Physics of the Earth and Planetary Interiors* 268: 35–53.

Faccenna, C., Di Giuseppe, E., Funiciello, F., Lallemand S., van Hunen J., (2009). Control of seafloor aging on the migration of the Izu–Bonin–Mariana trench. *Earth and Planetary Science Letters* 288: 386–398.

Freymuth, H., Vils, F., Willbold, M., Taylor, R. N., Elliott, T., (2015). Molybdenum mobility and isotopic fractionation during subduction at the Mariana arc. *Earth and Planetary Science Letters* 432: 176–186.

Čížková, H., Bin, C.R., (2015). Geodynamics of trench advance: Insights from a Philippine-Sea-style geometry. *Earth and Planetary Science Letters* 430: 408–415.

Griffiths, R.W., Hackney, R., van der Hilst, R.D., (1995). A laboratory investigation of effects of trench migration on the descent of subducted slabs. *Earth and Planetary Science Letters* 133: 1–17.

Samowitz, I.R., Forsyth, D.W., (1981). Double seismic zone beneath the Mariana Island arc. *Journal of Geophysical Research* 86: 7013–7021.

Miller, M., Kennett, B., Lister, G., (2004). Imaging changes in morphology, geometry, and physical properties of the subducting Pacific plate along the Izu–Bonin–Mariana arc. *Earth and Planetary Science Letters* 224: 363–370

Fujioka, K., Okino, K., Kanamatsu, T., Ohara, Y., (2002). Morphology and origin of the Challenger Deep in the Southern Mariana Trench. *Geophysical Research Letters* 29 (10): 1372.

Funiciello, F., Faccenna, C., Heuret, A., Lallemand, S., Di Giuseppe, E., Becker, T.W., (2008). Trench migration, net rotation and slab–mantle coupling. *Earth and Planetary Science Letters* 271: 233–240.

Heuret, A., Conrad, C.P., Funiciello, F., (2012). Relation between subduction megathrust earthquakes, trench sediment thickness and upper plate strain. *Geophysical Research Letters* 39(5): 131-138